\documentstyle[aps,twocolumn,epsf]{revtex}

\begin{document}
 
\title{Dilute Bose gas in a torus: vortices and persistent currents}
 
\author{D.S. Rokhsar}
\address{Department of Physics, 
University of California, Berkeley, CA 94720-7300}
 
\address{\em (September 18, 1997)}
 
\address{ {\em \bigskip \begin{quote}
We consider the excitation spectrum and stability of quantized vortices
in a weakly-interacting Bose gas trapped in a toroidal container, and 
discuss the driven rotation of such a condensate.
\end{quote} } }

%\pacs{03.75.Fi, 05.30.Jp, 32.80.Pj}

\maketitle

Quantization of circulation and the related phenomenon of persistent currents 
are two of the most striking properties of superfluids. \cite{Leggett}  The 
recent creation of atomic Bose-Einstein condensates \cite{BEC,Lithium} 
has renewed interest in the superfluidity of confined Bose systems. 
\cite{BP,Stringari,Burnett,Rokhsar-1,other-rotations,Java}
Here we discuss two related issues that pertain to the demonstration 
of these phenomena in trapped gases -- vortex stability and the driven 
rotation of condensates. 

Harmonically trapped Bose gases (and, more generally, gases confined 
to monotonically increasing potentials) cannot support stable vortices 
the absence of an externally imposed rotation. \cite{Rokhsar-1}  
In such a potential the vortex core is not pinned, and can slip out 
of the system, leaving behind a non-rotating condensate.\cite{Putter}
The simplest trap that could pin a vortex core is a toroidal
container, which confines the gas to a non-simply connected volume.  For 
theoretical ease, we consider a dilute Bose gas in a narrow torus whose
cross-sectional area ${A}$ is small compared with the square of its radius 
$R$.  This limit allows us to neglect transverse motion and 
to focus our attention on the more interesting azimuthal direction.
We regard the system as a ring \cite{Bloch} of radius $R$, and label
single-particle states by their azimuthal angular momentum.  We use 
the Bogoliubov approximation \cite{Bogoliubov} to explicitly calculate the 
properties of the dilute Bose gas in this geometry. 

Consider a collection of $N$ bosons of mass $M$ interacting via a 
$\delta$-function pseudopotential of strength $U = 4 \pi \hbar^2 a/M$, 
where $a$ is the $s$-wave scattering length.  The Hamiltonian is
\begin{eqnarray}
{\cal H} = \sum_q (\epsilon_q - \mu) b_q^\dagger b_q + 
\frac{U}{2V} \sum_{ \{ \stackrel{q_1 + q_2 =}{ q_3 + q_4} \} }
b_{q_1}^\dagger b_{q_2}^\dagger b_{q_4} b_{q_3} ,
\label{Hamiltonian}
\end{eqnarray}
where $\epsilon_q=$ $\hbar\omega q^2$ is the energy of the single-particle
state of angular momentum $q \hbar$ and 
$V = 2 \pi R A$ is the volume of the annulus.  The characteristic 
frequency $\omega = \hbar/(2MR^2)$ corresponds to the energy difference 
between the ground and first excited state of a single particle on the ring.

First, we discuss the stability of a Bose gas prepared in a quantized 
vortex of circulation $m(h/M)$, that is, a Bose condensate whose condensate 
wavefunction varies as $e^{im\phi}$. In the presence of a weak, static, 
asymmetric perturbation, angular momentum is not conserved.
In a rotating {\em normal} fluid, scattering of particles against such an 
asymmetric perturbation will ultimately bring the fluid to rest.
A characteristic feature of superfluids, however, is the absence of 
accessible final states into which the vortex can decay. \cite{Landau-crit}  
We show that unlike the case of a monotonic potential\cite{Rokhsar-1}, the 
pinning provided by the hole in the torus stabilizes the vortex if the
scattering length is sufficiently large and positive.

Second, we examine the production of a vortex starting from 
an initially non-rotating Bose condensate.   We discuss the response of
an $m=0$ condensate to a weak time-dependent potential that drives the
system to rotate at angular velocity $\Omega$.
At small $\Omega$, a weak, adiabatic perturbation imparts angular momentum 
to the condensate, but the corresponding moment of inertia 
$I \equiv \langle L_z \rangle / \Omega$ is proportional 
to the square of the perturbation strength and can be small compared with
$I_{\rm classical} = N M R^2$.  
This phenomenon of non-classical moment of inertia \cite{Leggett} 
is essentially the ``Meissner effect'' for superfluids, which ``expel'' 
rotation in the same way that superconductors expel magnetic fields.  
We determine the critical angular velocity in the toroidal geometry, which 
is simply related to the Landau critical velocity, and discuss the creation 
of vortices when this critical velocity is exceeded.

{\bf Excitation spectrum of a vortex.}
We begin by determining the excitation spectrum of a quantized vortex in
a dilute Bose gas, in the lab frame.  Although this spectrum can be derived 
by transforming to and from the co-rotating frame of reference, 
\cite{transform} it is instructive to calculate it directly in the lab frame 
using the Bogoliubov approximation. \cite{Bogoliubov,GP}  Following the
familiar protocol, we replace the creation and annihilation operators for 
the vortex condensate by a c-number, and retain only quadratic terms in the 
remaining creation and annihilation operators in eq. (\ref{Hamiltonian}).  
On a ring the vortex condensate of quantum number $m$
is the unique state $e^{im\theta}$. \cite{other-vortex}
The Bogoliubov Hamiltonian becomes
\begin{eqnarray}
{\cal H}_{\rm Bogo} &=& N E_C(m) +
\sum_{q \neq 0} [\epsilon_{m+q} - \epsilon_m + Un] 
b_{m+q}^\dagger b_{m+q} 
\nonumber \\
&+& \frac{Un}{2} \sum_{q \neq 0}
[ b_{m+q}^\dagger b_{m-q}^\dagger +  b_{m-q} b_{m+q} ] ,
\label{H-BOGO}
\end{eqnarray}
where $E_C(m) = \epsilon_m + Un/2$ is the Hartree-Fock energy per particle,
$n=N/V$ is the number density, and we have substituted the appropriate
chemical potential $\mu_m = \epsilon_m + Un$.  

The Hamiltonian (\ref{H-BOGO}) is easily solved by a canonical 
transformation to the quasiparticle operators
\begin{equation}
\beta_q^\dagger = u_q b_{m+q}^\dagger - v_q b_{m-q} ,
\label{canonical}
\end{equation}
where the amplitudes $u_q$ and $v_q$ must satisfy the normalization condition
$|u_q|^2 - |v_q|^2 = +1$ to ensure that Bose commutation relations
$[\beta_q, \beta_q^\dagger] = 1$ are preserved.  Note that excitations
are labeled by the angular momentum {\em relative to the condensate},
and are {\em not} states of definite angular momentum in the lab frame,
\cite{GP,Burnett}
due to anomalous scattering against the rotating condensate. There is a 
single quasiparticle excitation for each non-zero 
({\em i.e.,} positive and negative) integer $q$. 

We seek the amplitudes $u_q$, $v_q$ such that $\beta_q^\dagger$ is an 
eigenoperator of ${\cal H}_{\rm Bogo}$ with excitation energy $\hbar\omega_q$, 
{\em i.e.,} $[ {\cal H}_{\rm Bogo}, \beta_q^\dagger ] =$
$\hbar \omega_q \beta_q^\dagger$.
This leads to the (non-Hermitian) eigenvalue equation
\begin{eqnarray}
\left[
\begin{array}{cc}
{[\epsilon_{m+q} - \epsilon_m + Un]} & {Un} \\
{-Un} & {-[\epsilon_{m-q} - \epsilon_m + Un]} 
\end{array}
\right]
\left[
\begin{array}{c}
u_q \\
v_q
\end{array}
\right] 
\nonumber \\
= \hbar \omega_q 
\left[
\begin{array}{c}
u_q \\
v_q
\end{array}
\right] ,
\label{matrix-Bogo}
\end{eqnarray}
which has the two eigenvalues
\begin{equation}
\hbar \omega_q^{(\pm)} = \frac{\epsilon_{m+q}-\epsilon_{m+q}}{2} \pm
\sqrt{ e(m,q)^2 + 2 e(m,q) UN } ,
\label{om-1}
\end{equation}
where $e(m,q) \equiv$
$\frac{1}{2} [ \epsilon_{m+q} + \epsilon_{m-q} - 2\epsilon_m ].$ 
For particles on a ring (but not, for example, in a harmonic potential), 
$\epsilon_q = \hbar\omega q^2$, and $e(m,q) = \epsilon_q$. 

Only the positive sign in eq. (\ref{om-1})
permits the normalization $|u_q|^2 - |v_q|^2 = +1$.   The eigenvector
corresponding to the negative sign has $|v_q| > |u_q|$,
and cannot be normalized in this manner; it therefore cannot represent
a quasiparticle excitation.\cite{limit}
This unnormalizable solution instead represents
the quasi{\em hole} operator, $\beta_{-q}$, which is the
Hermitian conjugate of the (normalizable) operator $\beta_{-q}^\dagger$
that creates a quasiparticle of relative angular momentum $-q\hbar$. 
From eq. (\ref{matrix-Bogo}),  $u_{-q} = u_{q}$, and 
$v_{-q} = v_q$; for a ring, these coefficients are independent 
of $m$. 
 
We conclude that in the lab frame the energy of a quasiparticle 
of relative angular momentum $q\hbar$ above a vortex of 
quantum number $m$ is
\begin{eqnarray}
\hbar \omega_q^{(m)} = 2\hbar\omega m q + 
\sqrt{2Un \epsilon_q + (\epsilon_q)^2}.
\label{omega-q} 
\end{eqnarray}
In the strongly interacting limit $Un/\hbar\omega \gg 1$,
\begin{equation}
\hbar\omega_q^{(m)} \approx 2 \hbar\omega mq + \sqrt{2Un\hbar\omega} | q| 
+ O(q^2).
\label{large-U-omega}
\end{equation}
The frequencies $\omega_{+q}$ and $\omega_{-q}$ are different in a rotating 
condensate, reflecting the broken time-reversal and parity symmetries of the 
vortex.  This implies a ``Sagnac effect'' \cite{MAndrews} -- if we create 
clockwise and
counterclockwise propagating density fluctuations, they will move at
different velocities in the lab frame as they are carried along by 
the superflow.  If the $q=\pm 1$ modes are excited coherently, the 
resulting density fluctuation will precess at angular velocity
$[\omega_{1} - \omega_{-1}]/2 = 2 \omega m$, which is reminiscent of Vinen's
original detection of quantized vortices.

{\bf Persistent currents.}
We may use the spectrum (\ref{omega-q}) to assess the stability of a
vortex of quantum number $m$ in the presence of a static, weakly asymmetric 
perturbation.   We assume that the gas has been prepared in a vortex 
state at low temperatures.
A necessary condition for persistent currents is that the excitation 
spectrum of the current-carrying state must be positive for all relative 
angular momenta $q$: otherwise, even a weak asymmetric perturbation could 
excite a macroscopic number of such excitations, and the condensate would be 
destroyed.  In the non-interacting limit $Un \rightarrow 0$, 
vortices are {\em always} unstable in {\em any} geometry, since the total
energy of the system can 
always be reduced by transferring particles from the rotating condensate 
to the (non-rotating) single particle ground state.
Thus vortex stability depends critically on interparticle interactions.

A vortex of angular momentum $m > 0$ will be stable if $\omega_q$ is positive 
for all $q$, which ensures that excitations will not proliferate. It is easy
to see that if $\omega_{q=-1}$ is positive, then all other $\omega_q$ will
also be positive.  A little algebra then establishes the stability condition 
\begin{equation}
\frac{Un}{\hbar\omega} > 2m^2 - \frac{1}{2}  .
\label{persistent-condition}
\end{equation}
Evidently the dimensionless ratio of the mean-field interaction to the level
spacing must be sufficiently large for vortices to be stable.
This requirement reflects the fact that along the decay path of the vortex 
there must be an intermediate state with a node along the ring at which 
the winding number of the phase ``slips'' by $2\pi$, and the vortex ``core''
passes out of the torus. \cite{Leggett} The energetic cost
of introducing this density fluctuation depends on the 
compressibility of the gas; for sufficiently strong repulsion there is a 
barrier to the phase slip, and the vortex becomes stable.

{\bf Negative scattering length.}
It is interesting to consider the case of negative scattering length,
as is appropriate for ${}^7$Li, where Bose condensation of a limited number 
of atoms in confined geometries has been demonstrated.\cite{Lithium}   
The $m=0$ condensate is stable with respect to quasiparticle 
excitations as long as the argument of the square root in (\ref{omega-q}) is 
positive, {\em i.e.,} when interactions are sufficiently weak compared with 
the level spacing:
\begin{equation}
\frac{|U|n}{\hbar\omega} < \frac{1}{2} .
\label{pc-2}
\end{equation}
This is simply eq. (\ref{persistent-condition}), with $m=0$ and $U < 0$.  
When eq. (\ref{persistent-condition}) is violated, $\omega_q$ becomes 
complex, signaling an exponentially unstable mode.
A similar requirement holds for the harmonically trapped gas
\cite{negative-U}.

Similarly, we find that gases with negative scattering length 
in a toroidal geometry can never support persistent currents, since
vortices are then {\em always} unstable: $\omega_{q=-1}$ is always either
negative or complex when $U < 0$.  Again, this result follows from eq. 
(\ref{persistent-condition}), with $m \neq 0$.

{\bf Making vortices.}
A vortex state can be created by cooling a Bose gas through its condensation 
temperature while steadily ``stirring'' it at sufficiently large angular 
velocity $\Omega$, which can be accomplished with a rotating perturbation 
$V(\theta - \Omega t)$.  
In the limit of a weak perturbation, single-particle states can still
be adequately labeled by angular momentum.  In a steadily driven state
the macroscopically occupied single-particle wavefunction
will be that which minimizes the single-particle energy in the 
co-rotating frame,  $\epsilon_m^\prime = \epsilon_m - \hbar\Omega m$.  
For example, when $\omega < \Omega < 3\omega$ the $m=1$ vortex will be 
formed at the Bose condensation temperature. 
When the rotating perturbation is turned off, a vortex in a toroidal
trap will be stable if eq. (\ref{persistent-condition}) is satisfied. 

{\bf Landau criterion.}
Can a vortex be generated by stirring an initially non-rotating condensate?
To simplify the discussion we assume that the perturbation is weak, and 
consider initial temperatures well below $T_c$ so that we may continue to
use the Bogoliubov approximation.  The time-dependent rotating perturbation 
becomes
\begin{equation}
{\cal H}_{\rm pert} = \sum_{q \neq 0} \sqrt{N} 
\tilde{V}_q (u_q+v_q) e^{-i\Omega q t}
(\beta_q^\dagger + \beta_{-q}),
\label{H-pert}
\end{equation}
where $\tilde{V}_q$ is
the Fourier transform of the perturbation $V(\theta)$, and we take 
$\Omega > 0$.  
% For clarity, we specifically consider the case of the 
% perturbation $V(\theta) = 2V \cos(\theta)$, {\em i.e.,} a rotating bump.

There are two simple limits to consider.  First, we may consider turning
on the rotating perturbation adiabatically, {\em i.e.,} 
$V(\theta,t) = [1 - e^{-t/\tau}] V(\theta - \Omega t),$
where the ``turning on'' time $\tau$ is long compared with 
$|\omega-\Omega|^{-1}.$  Inverting the Bogoliubov transformation
(\ref{canonical}), we see that the condensate becomes
\begin{eqnarray}
\Psi(\theta,t) &\equiv& \langle \hat{\psi}(\theta) \rangle 
= \frac{1}{\sqrt{2\pi N}} \{ \langle b_{q=0} \rangle +
\sum_{q \neq 0} e^{iq\theta} \langle b_q \rangle \}
\\
&=& \frac{1}{\sqrt{2\pi N}}
\{ \sqrt{N} + \sum_{q \neq 0} e^{iq\theta}
[u_q \langle \beta_q \rangle + v_q \langle \beta_q^\dagger \rangle] 
\} ,
\label{adiabatic-psi} 
\end{eqnarray}
where by first order time-dependent perturbation theory
\begin{equation}
\langle \beta_q \rangle = \langle \beta_q^\dagger \rangle^* =
\frac{\sqrt{N} \tilde{V}_q (u_q + v_q)}{\hbar(\omega_q - \Omega q)} 
e^{-i\Omega q t} 
\label{expect}
\end{equation}
when the perturbation is fully developed ({\em i.e.,} $t \gg \tau$).

The condensate responds at the driving frequency $\Omega$ by forming
density minima and maxima that track the peaks and valleys of the
rotating perturbation.  (This is easily seen by  considering 
$\Psi(\theta-\Omega t)$, {\em i.e.,} in the co-rotating frame.)
If $V(\theta) = 2 V_q cos(q\theta)$, then the steady state
$\Psi(\theta,t\gg\tau)$ has non-zero angular momentum
\begin{equation}
\langle L_z \rangle = q N\hbar 
% \sum_{q > 0}
\frac{ 4\omega_q\Omega |\tilde{V}_q|^2 (u_q + v_q)^2}
{\hbar^2 (\omega_q^2 - q^2\Omega^2)^2} ,
\label{ang-mo-psi}
\end{equation}
but $\Psi$ only describes a vortex when its phase winds 
around by $2\pi$, which requires that the coefficient of $e^{i\theta}$ be
large compared to that of $1$ and $e^{-i\theta}$.  For a weak perturbation
this requires the near resonance condition $\Omega \sim \omega_{q}/q$.  
The energy transferred to the gas per unit angular momentum,
$[\langle H \rangle - E_G]/\langle L_z \rangle$, is 
$(\omega_q^2 + \Omega^2)/2q\Omega$, which tends to $\omega_q/q$ near resonance.

If the same perturbation is turned on suddenly, it induces
transitions between states that differ by energy 
$\hbar \Omega q$ and by angular momentum $\hbar q$. This resonance
condition requires $\Omega = \omega_q/q$ for some $q$. 
For the Bogoliubov spectrum (\ref{omega-q}) with $m=0$, the minimum value
of $\omega_q/q$ occurs for $q=+1$, and the critical angular 
velocity is 
\begin{equation}
\Omega_{\rm critical} = \omega_{q=+1}^{(m=0)} = \omega
\sqrt{1 + 2 \left[ \frac{Un}{\hbar\omega}\right] }.
\label{critical-Omega}
\end{equation}
In the limit of large $Un/\hbar\omega$, 
$\Omega_{\rm critical} \rightarrow$ $[{Un}/({MR^2})]^{1/2} .$

Eq. (\ref{critical-Omega}) is nothing but the famous Landau criterion 
\cite{Landau-crit} applied to the dilute gas in a rotating torus. 
For a dilute gas, there is no roton minimum, and the Landau critical
velocity is the speed of sound $c = [Un/M]^{1/2}$.
The velocity $v_{\rm pert}$ of the rotating perturbation relative to 
the (non-rotating) condensate is  $\Omega R$, so according to Landau's  
argument the driven Bose system will be stable with respect to the 
proliferation of elementary excitations as long as $\Omega R$ does not 
exceed $c$, which recovers (\ref{critical-Omega}) in the large 
$Un/\hbar\omega$ limit.

{\bf The driven gas.}
When the gas is driven at or near resonance, its angular momentum and 
energy both grow beyond the linear response regime.  To understand the 
fate of the driven gas, it is useful to first consider the (many-body)
ground state 
of a Bose gas subject to the constraint that its total angular momentum 
is $M_z \hbar$.  The {\em overall} ground state is always at $M_z = 0$, 
since the true ground state must be nodeless; for small $M_z$ where
quasiparticle interactions may be neglected,
$E_G(M_z) = E_G(0) + \hbar\omega_{q=1}|M_z|$.

For sufficiently large $Un/\hbar\omega$, the ground state energy 
$E_G(M_z)$ will have more than one local minimum.  A full calculation 
of $E_G(M_z)$ is beyond the scope of this paper, but we may understand its 
general features by considering the family of condensates of the form
\begin{equation}
\Psi_x(\theta) = \frac{1}{2\pi} [ \sqrt{1-x} + \sqrt{x} e^{i\theta} ],
\label{vary-psi}
\end{equation}
where $0 \leq x \leq 1$.  The total angular momentum of (\ref{vary-psi})
is $\langle M_z\rangle = Nx$, and its total (Gross-Pitaevskii) energy is 
$E_G(M_z)= \hbar\omega M_z + \frac{1}{2}Un[N +2 M_z - 2 M_z^2/N]$,
which develops a second local minimum at $M_z = N$ for 
$Un/\hbar\omega > 1$.  
The barrier between the minima at $M_z=0$ and $N$ arises from the presence 
of a node in $\Psi$ when the winding number of the phase changes by $2\pi$, 
which occurs at $x = 1/2$ in eq. (\ref{vary-psi}).  The density fluctuation
that accompanies the node at intermediate $x$ costs energy due to the
compressibility of the interacting gas; for small and large $x$, however,
the magnitude of $\Psi$ is nearly uniform. and no additional energy costs
arise.  Since $\partial^2 E_G/\partial M_z^2 < 0$ for $0 < M_z < N$, only the 
$M_z=0$ or $N$ states can be thermodynamically stable.

Armed with this understanding of $E_G(M_z)$ we can develop a scenario
for spinning a stationary condensate into a vortex. 
First, we imagine an adiabatic perturbation which increases the angular 
momentum of the gas while keeping it in the corresponding ground state.  
Then during a short time interval the energy increment $\delta E$ and 
the angular momentum increment $\delta M_z$ must be related by
$\delta E = [d E_G/d M_z] \delta M_z$,
which requires that the angular frequency of the perturbation vary as
$\Omega(t) \equiv [d E_G/d M_z]|_{M_z(t)}$. 
(The absolute values of $\delta M_z$ and $\delta E$ depend on the strength
of the perturbation.)
Since $E_G(M_z)$ has a maximum, the angular velocity $\Omega$ 
diminishes as the total angular momentum increases. 
For our simple model, $\Omega (M_z) = \omega + Un - 2Un(M_z/N)$.  
% The decrease in frequency can be viewed as a reduction in quasiparticle
% excitation energy as the condensate distorts.

Eventually, the maximum of $E_G(M_z)$ is reached.  If the angular momentum 
is increased further (by continuing to drive the gas at low frequency), the 
gas enters a regime in which $\partial E_G/\partial M_z$ becomes negative,
and the ground state energy {\em decreases} with further increase in $M_z$.
Since the gas is thermodynamically unstable, in the absence of any further
drive it will relax towards the nearest free energy minimum, which is now 
the vortex condensate with $m=1$.  The energy liberated in this relaxation
process goes into generating excitations above the ground state, and heating 
this normal fluid.  
The angular momentum gained by the condensate appears at the expense of 
an opposite angular momentum change in the normal fluid, which ultimately 
decays due to scattering by small inhomogeneities in the trap.

A back-of-the-envelope estimate suggests that heating due to non-adiabatic 
driving of the condensate is not a problem.
If we simply drive the gas at frequency $\sim \Omega_{\rm critical}$ long 
enough for it to acquire angular momentum $N\hbar$, then it will absorb an 
energy of order $\sim N\hbar\Omega_{\rm critical}$. 
At this point we rapidly turn off the perturbation (or turn its 
frequency down to $\omega < \Omega < 3 \omega$ as discussed above), and 
allow the gas to reach a new equilibrium.  If the gas 
has not heated up past $T_c$, then it should relax into the $m=1$ vortex.
The condition is that the total energy transferred during the excitation 
period, $\sim N\hbar\Omega_{\rm critical}$, must be less than
$\sim N k_B T_c + E_G(M_z=N)$.  
Taking $k_B T_c \sim \hbar^2 n^{2/3}/M$ as appropriate for
a dilute gas of density $n$, we find that a vortex can be created in this
manner if $n > [a^3/R^6]$.  If $R \gg a$, there is no conflict with the 
dilute gas condition $na^3 \ll 1$.

{\bf Acknowledgements.}
I thank M. Andrews, K. Burnett, E. Cornell, J.C. Davis, E. Hill, M. Jebrial,
D. Jin, S. Kivelson, W. Ketterle, R. Packard, and K. Singh for useful 
conversations.  This work was supported by NSF DMR 91-57414 and by the
Miller Institute for Basic Research in Science.

\begin{equation}
E_{GP} [ \Psi ] \equiv \int d{\bf r} 
[ \Psi^* H_o \Psi + \frac{UN}{2} |\Psi|^4 ].
\label{G-P}
\end{equation}
Consider the energy of a condensate $\Psi + \phi$, where 
$\delta E_{GP}/\delta \Psi = \mu \Psi$ and $\phi$ 
is a small (complex) variation.  To quadratic order in $\phi$,
the variation in $E_{GP}$ is 
\begin{equation}
\int d{\bf r} \left[ \phi^* [H_o - \mu] \phi + \frac{Un}{2}
[\Psi^2 (\phi^*)^2 + (\Psi^*)^2 \phi^2 + 4 |\Psi|^2|\phi|^2 ] \right].
\end{equation}
For a vortex on a ring, $\Psi = e^{im\theta}$, and this
quadratic form can be partially diagonalized by defining
\begin{equation}
\phi_q(\theta) = a_q e^{i(m+q)\theta} + b_q^* e^{i(m-q)\theta}
\label{partial-diag}
\end{equation}
where ($a_q, b_q$) is the eigenvector of the (Hermitian) matrix 
\begin{eqnarray}
\left[
\begin{array}{cc}
{[\epsilon_{m+q} - \epsilon_m + Un]} & {Un} \\
{Un} & {[\epsilon_{m-q} - \epsilon_m + Un]} 
\end{array}
\right] .
\label{matrix-stability}
\end{eqnarray}
Note that (\ref{partial-diag}) and (\ref{matrix-stability}) are related 
to but distinct from (\ref{canonical}) and (\ref{matrix-Bogo}).
The vortex is then locally stable if the two eigenvalues
\begin{equation}
\lambda_q^{(\pm)} = \epsilon_{q}+ Un \pm
[ (2m^2 \epsilon_q)^2 + (Un)^2 ]^{1/2}
\label{stab}
\end{equation}
are both positive for all $q$.  This condition is the same as 
(\ref{persistent-condition}).  More generally, for arbitrary traps one
finds that the positivity of the elementary excitation spectrum coincides
with the stability of the Gross-Pitaevskii energy.

% \bibitem{Landau}{E.M. Lifshitz and L.P. Pitaevskii, 
% {\em Statistical Physics} (Pergamon Press, Oxford, 1980).}


\begin{references}
\bibitem{other-vortex}
For other traps one must first carry out the
non-trivial step of determining the radial and axial dependence of the
vortex condensate, as in refs. \onlinecite{GP,Stringari}.
\bibitem{limit}
The appropriateness of the positive root is confirmed by considering
the non-interacting limit $\omega_q(Un \rightarrow 0) =$ 
$\epsilon_{m+q} - \epsilon_m$, where we 
recover the result that the elementary excitation of relative 
angular momentum $q$ corresponds to the transfer of a particle 
from the vortex condensate to the single-particle state with angular 
momentum $m+q.$  The negative root of (\ref{om-1}) cannot be identified
with an excitation of the ground state in this limit.
\bibitem{stability}
A condensate $\Psi$ will be stable if it is a local minimum
of the Gross-Pitaevskii energy functional
\end{references}
\end{document}